\documentclass[10pt,twocolumn,letterpaper]{article}

\usepackage[pagenumbers]{iccv} %

\newcommand{\red}[1]{#1}

\usepackage[accsupp]{axessibility}
\usepackage[T1]{fontenc}

\newcommand{\name}{MeshPad}
\newcommand{\best}[1]{\textbf{#1}}
\newcommand{\second}[1]{#1}

\newcommand{\para}[1]{\vspace{0.1cm}\noindent\textbf{#1}}

\usepackage{multirow}
\usepackage{colortbl}
\usepackage{nicefrac}

\definecolor{seq_green}{RGB}{102, 169, 157}
\definecolor{seq_purple}{RGB}{168, 102, 147}

\makeatletter
\newcommand{\defaultautoref}[2]{%
  \@ifundefined{r@#1}{#2}{\autoref{#1}}%
}
\makeatother

\expandafter\def\expandafter\normalsize\expandafter{%
    \normalsize%
    \setlength\abovedisplayskip{4pt}%
    \setlength\belowdisplayskip{4pt}%
    \setlength\abovedisplayshortskip{2pt}%
    \setlength\belowdisplayshortskip{2pt}%
}

\definecolor{iccvblue}{rgb}{0.21,0.49,0.74}
\usepackage[pagebackref,breaklinks,colorlinks,allcolors=iccvblue]{hyperref}

\title{\name{}: Interactive Sketch-Conditioned Artist-Reminiscent \\Mesh Generation and Editing}

\author{ Haoxuan Li$^1$ \quad Ziya Erkoc$^1$ \quad Lei Li$^1$ \quad Daniele Sirigatti$^2$ \quad Vladislav Rosov$^2$ \\ Angela Dai$^1$ \quad Matthias Nie{\ss}ner$^1$ \vspace{0.15cm} \\
     {
        \begin{tabular}[t]{c c}%
          $^{1}$Technical University of Munich &
          $^{2}$AUDI AG \\
        \end{tabular}
     }  
}

\begin{document}

\twocolumn[{%
\renewcommand\twocolumn[1][]{#1}%
\maketitle
\includegraphics[width=\linewidth]{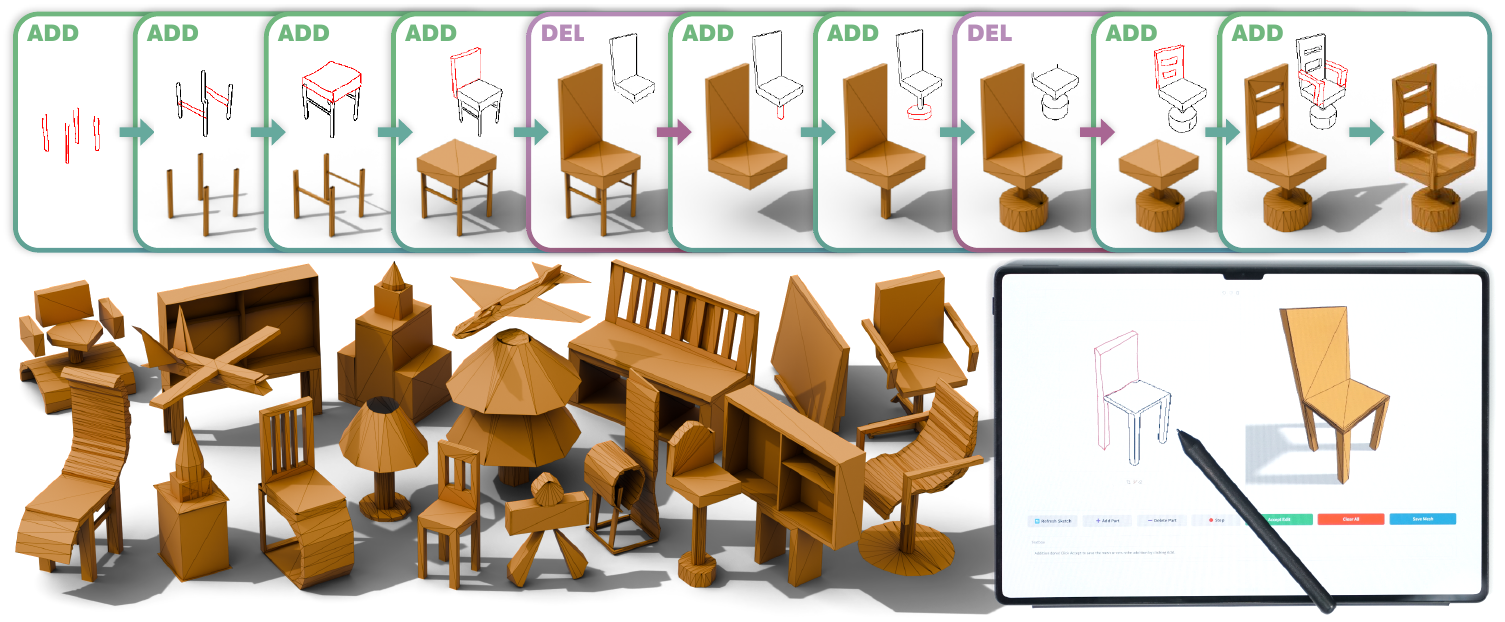}
\vspace{-0.6cm}
\captionof{figure}{\name{} enables interactive mesh creation and editing with sketches. 
We decompose this complex task into two sketch-conditioned operations:
{\color{seq_green}addition} and {\color{seq_purple}deletion}.
\textbf{Top}: our method allows for a user to create and modify artist-reminiscent triangle meshes by simply drawing and editing 2D sketches, achieving intuitive and interactive 3D modeling. \textbf{Bottom (left)}: our method generates a variety of complex yet compact meshes. \textbf{Bottom (right)}: our interactive user interface allows users to iteratively edit the mesh, with each edit step taking a few seconds.\vspace{1em}}
\label{fig:teaser}
}]

\let\thefootnote\relax\footnotetext{\vspace{-0.3cm}Project page: \href{https://derkleineli.github.io/meshpad/}{https://derkleineli.github.io/meshpad/}}
\begin{abstract}
We introduce \name{}, a generative approach that creates 3D meshes from sketch inputs. 
Building on recent advances in artist-reminiscent triangle mesh generation, our approach addresses the need for interactive mesh creation. 
To this end, we focus on enabling consistent edits by decomposing editing into `deletion' of regions of a mesh, followed by `addition' of new mesh geometry.
Both operations are invoked by simple user edits of a sketch image, facilitating  an iterative content creation process and enabling the construction of complex 3D meshes.
Our approach is based on a triangle sequence-based mesh representation, exploiting a large Transformer model for mesh triangle addition and deletion.
In order to perform edits interactively, we introduce a vertex-aligned speculative prediction strategy on top of our additive mesh generator. 
This speculator predicts multiple output tokens corresponding to a vertex, thus significantly reducing the computational cost of inference and accelerating the editing process, making it possible to execute each editing step in only a few seconds.
Comprehensive experiments demonstrate that \name{} outperforms state-of-the-art sketch-conditioned mesh generation methods, 
achieving more than $22\%$ mesh quality improvement in Chamfer distance,
and being preferred by $90\%$ of participants in perceptual evaluations.
\end{abstract}
    
\vspace{-0.5cm}
\section{Introduction}
\label{sec:intro}

\begin{figure*}[ht]
    \centering
    \includegraphics[width=\linewidth]{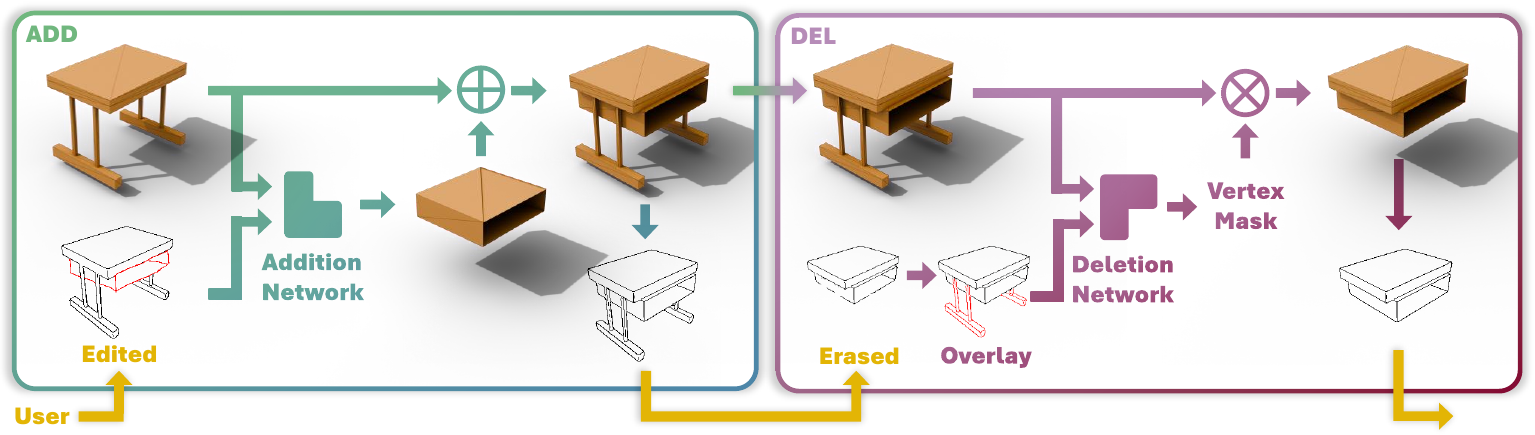}
    \vspace{-0.8cm}
    \caption{Method Overview. 
    We show the decomposition of mesh creation and editing into {\color{seq_green}addition} and {\color{seq_purple}deletion} operations (left and right, respectively).
    \textbf{\color{seq_green}ADD}: in mesh addition, a transformer generates new mesh regions  corresponding to newly added strokes (in red) in the input sketch. The generated mesh triangles are then merged with the existing mesh. \textbf{\color{seq_purple}DEL}: in mesh deletion, we show a deletion operation applied to the mesh addition output. 
    We erase sketch strokes to remove corresponding mesh regions, and overlay the erased strokes in red alongside untouched regions to provide more context to the deletion network.
    We then predict which mesh vertices correspond to regions to be deleted, and prune the mesh accordingly. 
    After each addition or deletion operation, we automatically generate an updated sketch corresponding to the current output mesh, to enable further sketch-based editing.}
    \vspace{-0.2cm}
    \label{fig:overview}
\end{figure*}

Triangle meshes are one of the most predominant 3D representations used in 3D production applications, from video games to virtual reality and movies.
Mesh creation and editing is thus a central element of computer graphics.
In contrast to 3D representations such as voxels~\cite{dai2017shapecompletion, dai2020sgnn, ren2024xcube}, points~\cite{vahdat2022lion, zhou2021pvd, yang2019pointflow}, or neural implicit representations~\cite{hertz2022spaghetti, mildenhall2021nerf, kerbl20233d}, triangle meshes represent surfaces in a compact, structured fashion, as well as enabling efficient fine-scale detail and naturally integrating into modern rendering and editing  pipelines, achieving high fidelity with relatively few primitives. 

Recent advances in generative 3D models have shown significant potential in generating 3D meshes directly \cite{siddiqui2023meshgpt, chen2024meshanything, chen2024meshanythingv2artistcreatedmesh, tang2024edgerunner, hao2024meshtron}; however, these output meshes are not editable, which is a crucial component for artistic design in content creation. 
More specifically, artistic content creation is an iterative process that encompasses not only the initial generation, but also requires multiple cycles of manipulation and editing in order to refine an output to achieve a precise artistic vision. 
In order for such editing to fit with content creation pipelines, various edits must be performed interactively,
affecting only the intended region of the mesh.

We thus propose \name{}, an interactive sketch-based approach for 3D mesh generation and editing. 
From an input sketch of a shape drawn by a user, we produce a corresponding 3D mesh. 
The resulting mesh can then be edited simply by editing the sketch.
To achieve efficient and precise mesh editing, we decompose this into simpler subtasks: 
\textit{deletion} and \textit{addition} of mesh geometry by removing or adding strokes in the sketch.
Crucially, this modeling paradigm also leads to easy supervised training of both deletion and addition of elements in the mesh (by simply removing and adding back parts of ground-truth meshes), 
without requiring collection of real mesh editing sequences for supervision.
Our editing-based approach allows for the generation of more complex triangle meshes by iteratively applying a series of edits to construct the final shape.

Algorithmically, \name{} leverages a hybrid approach combining an autoregressive network for addition and a token-classification network for deletion. Unlike existing artist-reminiscent mesh generation methods, 
our addition network generates partial shapes only corresponding to new sketch strokes, rather than repeatedly synthesizing the entire mesh.
This naturally solves the problem of preserving unedited mesh regions during editing. 
Additionally, based on the triangle sequence representation~\cite{chen2024meshanythingv2artistcreatedmesh}, 
we introduce a vertex-aligned speculative decoder to accelerate autoregressive generation and reduce computational time.
Instead of predicting one vertex coordinate at a time, the speculator allows us to predict $3$ coordinates, i.e., one vertex, at a time. We find that jointly training the speculative head and aligning it with the vertex tokens reduces generation time while maintaining generation quality. 
\red{We further demonstrate our method through a user interface (\autoref{fig:teaser}) that allows users to interactively edit 3D meshes by drawing on 2D sketches.}
Extensive experiments show that our method outperforms state-of-the-art methods in sketch-conditioned mesh generation, 
producing cleaner meshes that more accurately align with the sketch input and excelling in partial mesh editing tasks.
Finally, our speculative prediction acceleration enables interactive mesh generation and editing within seconds while maintaining overall shape quality.

In summary, our contributions are:
\begin{itemize}
    \item We introduce a novel method for interactive mesh creation and editing by decomposing the process into  addition and deletion. This enables easy training without requiring collection of edited ground-truth meshes, and enables finer-grained control over an iterative 3D mesh creation process.
    \item Our vertex-aligned speculative classification head for addition notably accelerates the triangle mesh generation by $2.2\times$ without quality loss. 
\end{itemize}

\section{Related Work} 
\label{sec:relatedwork}

\para{3D Mesh Generation.}
Direct mesh generation offers the compelling advantage of producing outputs that closely resemble artist-crafted 3D shapes.
Early approaches proposed various parameterizations of an irregular mesh structure, including altases \cite{groueix2018papier} and graphs \cite{dai2019scan2mesh}.
Recently, Polygen~\cite{nash2020polygen} and MeshGPT~\cite{siddiqui2023meshgpt} have demonstrated the remarkable potential of transformers in autoregressively generating artist-crafted meshes.
Follow-ups such as MeshAnything~\cite{chen2024meshanything}, MeshAnything V2~\cite{chen2024meshanythingv2artistcreatedmesh}, EdgeRunner~\cite{tang2024edgerunner}, and Meshtron~\cite{hao2024meshtron} each proposed improvements in tokenization, attention mechanisms, or shape coverage, pushing the limit of direct mesh generation. PolyDiff~\cite{alliegro2023polydiffgenerating3dpolygonal} instead uses a diffusion backbone for polygonal mesh synthesis. 
Despite solid progress, these methods focus on full-shape generation and cannot easily handle local editing tasks.

\para{Conditional 3D Generation.}
The history of sketch-driven methods starts with techniques that inflate 2D contours \cite{igarashi2006teddy, nealen2007fibermesh} or retrieve 3D models from sketch-based queries \cite{eitz2012sketch, funkhouser2003search}. 
Modern sketch-based modeling methods explored representations like SDF \cite{zheng2023lasdiffusion, borth2024sketch2shape}, NeRF \cite{mikaeili2023sked}, and CAD commands \cite{li2022free2cad}.
With the advancement in large text-to-image models, DreamFusion~\cite{poole2023dreamfusion} proposes score distillation sampling with 2D diffusion for text-to-3D generation, followed by Magic3D~\cite{Lin_2023_CVPR}, Fantasia3D~\cite{chen2023fantasia3d}, and Meta 3D AssetGen~\cite{siddiqui2024assetgen}.
Text conditions lack precision for fine-grained control \cite{xu2023instructp2plearningedit3d} and additional image condition is often required \cite{qian2024pushing, bala2024edify}.
In contrast, our method uses iterative sketch conditioning for fine-grained shape control. 

\para{3D Shape Editing.}
With the same concept of SDS, many 3D shape editing methods rely on neural representations \cite{khalid2025latenteditor, liu2025maskeditor} and use region-specific masks \cite{Chen2024gaussianeditor, barda2024instant3dit, chen2024mvedit}. 
SPAGHETTI~\cite{hertz2022spaghetti}, SALAD~\cite{koo2023salad}, and PartGen~\cite{chen2024partgen} advance part-level neural shape editing with disentangled control, diffusion-based manipulation, and multi-view diffusion for generating and editing meaningful 3D parts, respectively.

Recently, sketch-based 3D editing has emerged as an intuitive alternative to text- and image-based methods. 
SKED~\cite{mikaeili2023sked} and SketchDream~\cite{liu2024sketchdream} leverage sketch-conditioned 2D diffusion models for 3D modeling. Other approaches, including SENS~\cite{Binninger:SENS:2024}, Doodle your 3D~\cite{bandyopadhyay2023doodle}, and Masked LRM~\cite{gao2024maskedlrm}, focus on abstraction-aware part manipulation, robust part isolation, and masked reconstruction for fast local edits, respectively.

However, such works struggle to produce an artist-reminiscent mesh. 
Techniques based on polygonal meshes or other explicit representations can rely on an expensive optimization \cite{aigermann2022jacobianfields, gao2023textdeformer}, deformation priors \cite{tang2022neural}, specialized parametric families \cite{elrefaie2024drivaernet} or procedural generators \cite{zhao2024di}. 
Such approaches do not generalize easily to arbitrary geometry and have difficulties making local edits without conflicting with the global constraints. 
In contrast, our approach addresses these issues using a sequence-based mesh representation to perform partial editing by removing and adding triangles.

\para{Speculative Decoding.}
Speculative decoding accelerates autoregressive generation by predicting multiple tokens in parallel~\cite{pmlr-v202-leviathan23a, chen2023accelerating, miao2023specinfer, wertheimer2024speculator}.
We extend this approach to mesh sequences with a vertex-aligned speculator, enabling faster partial additions for iterative mesh editing.

\section{Method}

\name{} enables interactive mesh creation and editing by iteratively performing addition and deletion operations guided by input sketches.
An input sketch image $\mathcal{I}$ guides the generation of a 3D mesh $\mathcal{M}$ as a sequence of mesh triangles. $\mathcal{M}$ can be further edited by removing strokes from $\mathcal{I}$ to delete mesh regions, or adding new strokes for new geometry.
To achieve interactive rates for editing, we introduce a vertex-aligned speculator into our mesh addition transformer to accelerate its autoregressive generation.
An overview of our approach is shown in \autoref{fig:overview}.

\subsection{Mesh Creation and Editing}
\label{sec:task}
We decouple sketch-based mesh creation and editing into two subtasks: addition and  deletion. 
Both operations are conditioned on a sketch image $\mathcal{I}$, 
a bitmap where colored line strokes indicate regions to be added or deleted.
We define a mesh $\mathcal{M}$ as a set of triangles $\mathcal{M}=\{\mathcal{F}\}$, where triangles
$\mathcal{F} = \{\boldsymbol{v}_1, \boldsymbol{v}_2, \boldsymbol{v}_3\}$ and $\boldsymbol{v}\in \mathbb{R}^3$ represents vertices.

Line strokes in the input sketch image $\mathcal{I}$ are divided into two mutually exclusive sets, $\mathcal{I}_k$ and $\mathcal{I}_r$, colored black and red in our visualizations, respectively. When performing an addition or deletion, $\mathcal{I}_r$ corresponds to the user edit, i.e., the part to be added or deleted, respectively, while $\mathcal{I}_k$ represents the untouched sketch regions. We denote the mesh parts corresponding to $\mathcal{I}_r$ and $\mathcal{I}_k$ as mutually exclusive sets $\mathcal{M}_r$ and $\mathcal{M}_k$, respectively, and thus $\mathcal{M}_r \cup \mathcal{M}_k = \mathcal{M}$.
The operation---addition or deletion---is 
determined by whether the $\mathcal{I}_r$ is added as new strokes (addition), or erased from existing strokes (deletion).
This also enables simple supervision of both deletion and addition networks by removing or adding back random regions of mesh geometry of 3D shapes, without requiring ground-truth sequences of mesh edits (see \autoref{sec:data_processing} for more detail).

\para{Sketch-Conditioned Mesh Deletion.}
Given a current mesh state $\mathcal{M}$ and a sketch image $\mathcal{I}$ containing a non-empty set of deletion strokes $\mathcal{I}_r$, the goal is to obtain a mesh $\mathcal{M}_k \subsetneq \mathcal{M}$ that excludes parts corresponding to $\mathcal{I}_r$. This is formulated as a binary classification task over the sequence of triangles composing $\mathcal{M}$; those predicted as corresponding to $\mathcal{I}_r$ are then removed.

We employ a transformer-based architecture (\autoref{subsec:arch}) to predict a binary label for each vertex in the mesh sequence, forming a set of vertices $\mathcal{V}^\prime_r$ corresponding to $\mathcal{I}_r$. Then we predict the target mesh $\mathcal{M}^\prime_k$ by deleting triangles with any vertex predicted as deleted:
\begin{equation}
    \mathcal{M}^\prime_r = \left\{\mathcal{F}\in \mathcal{M}|\exists \boldsymbol{v}\in\mathcal{F}: \boldsymbol{v} \in \mathcal{V}^\prime_r\right\};\quad \mathcal{M}^\prime_k=\mathcal{M}\setminus\mathcal{M}^\prime_r,
    \label{eq:deletion}
\end{equation}
where $\setminus$ is the set subtraction operator.

We train this by automatically generating 2D sketches and removal regions from 3D shapes in a self-supervised fashion (\autoref{sec:data_processing}) to obtain the ground truth $\mathcal{V}_r=\bigcup_{\mathcal{F}\in\mathcal{M}_r}\mathcal{F}$ and set the binary label of each vertex as:
\begin{equation}
    l_{\boldsymbol{v}} =
    \begin{cases} 
    0 & \text{if } \boldsymbol{v} \in \mathcal{V}_r, \\
    1 & \text{otherwise}.
    \end{cases}
    \label{eq:binary_deletion_label}
\end{equation}
The network is trained with a binary cross-entropy loss between the predicted binary probability per vertex and the vertex label. 
Once deletion is performed, we assign ${\mathcal{M} \gets \mathcal{M}^\prime_k}$ and ${\mathcal{M}_k\gets \mathcal{M}^\prime_k}$, and automatically generate the updated $\mathcal{I}_k$ 
 corresponding to the new output mesh $\mathcal{M}$ (\autoref{sec:data_processing}), which can be further edited.

\para{Sketch-Conditioned Mesh Addition.}
The addition task is the inverse process of deletion. The input is a mesh $\mathcal{M}_k$ and a sketch $\mathcal{I}$ containing non-empty $\mathcal{I}_r$, and the goal is to generate the mesh $\mathcal{M}$. We use a similar architecture as with deletion to autoregressively generate tokens of $\mathcal{M}^\prime_r$ and merge it with $\mathcal{M}_k$ to obtain the predicted mesh $\mathcal{M}^\prime$. This is supervised by a cross-entropy loss between the ground-truth mesh tokens of $\mathcal{M}_r$ and the predicted token probabilities. After merging, we assign ${\mathcal{M} \gets \mathcal{M}^\prime}$ and $ {\mathcal{M}_k \gets \mathcal{M}^\prime}$.

We automatically generate an updated sketch $\mathcal{I}_k$ reflecting the latest mesh geometry. 
This enables iterative mesh generation and editing. A user first draws a sketch to produce an initial mesh or loads in a mesh from which we automatically generate a corresponding sketch. 
Editing can then be performed on this initial mesh and sketch; a user can erase a sketch part with an eraser tool, labeling part of $\mathcal{I}_k$ as $\mathcal{I}_r$ and triggering deletion of the corresponding mesh part.
The user can also draw new strokes alongside the current $\mathcal{I}_k$ to introduce new mesh geometry.
This iterative editing process avoids re-synthesis of the whole shape, preserving mesh structures in unedited regions.
Note that mesh generation from scratch is treated as addition to an empty mesh.

\begin{figure}[t]
    \centering
    \includegraphics[width=\linewidth]{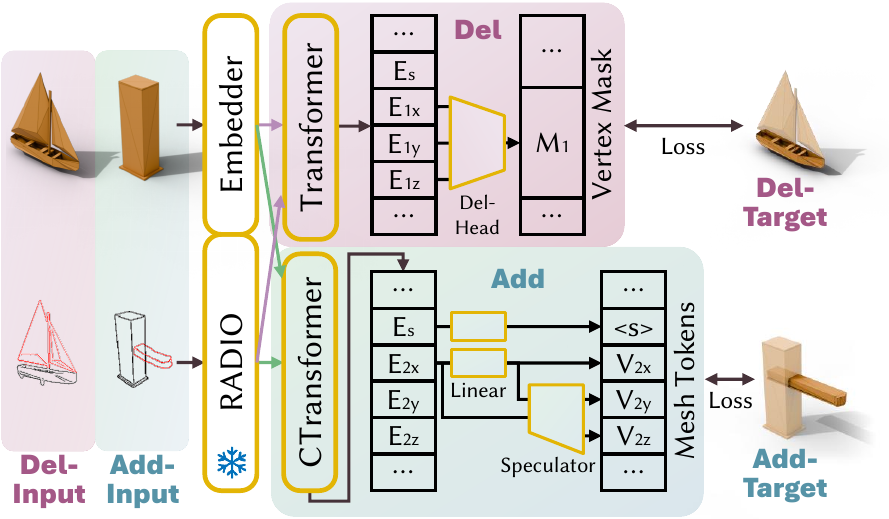}
    \vspace{-0.6cm}
    \caption{Model Architecture. We use the Open Pre-trained Transformer (OPT) in both the {\color{seq_green}addition} and {\color{seq_purple}deletion} networks as the backbone to process the embedded triangle sequence and the sketch embeddings from a pre-trained image foundation model, RADIO (frozen during training). \textbf{Top (right)}: the {\color{seq_purple}deletion} network is a classification network that labels each input mesh vertex for removal. The deletion head takes the $xyz$ coordinate embeddings of a vertex as input and produces a masking label for the vertex. \textbf{Bottom (right)}: the {\color{seq_green}addition} network is an autoregressive generation model that predicts new mesh tokens with a speculator. The \textit{CTransformer} backbone uses causal attention instead of full attention. During the speculative prediction, we align the speculator to the vertex axis so that it always takes the $x$ coordinate as input and predicts the vertex's $y$ and $z$ coordinates. $\text{E}$, $\text{M}$, $\text{V}$, and $\text{<s>}$ represent Transformer hidden states, vertex masks, vertex coordinate tokens, and the $\text{<split>}$ token, respectively. \vspace{-0.6cm}}
    \label{fig:network}
\end{figure}

\subsection{Network Architecture}\label{subsec:arch}
The detailed model structure is shown in \autoref{fig:network}. We use a pre-trained image encoder RADIO 2.5-h~\cite{Ranzinger2024radio} to encode the sketch input as tokens.
We follow MeshAnythingV2~\cite{chen2024meshanythingv2artistcreatedmesh} to use the same Open Pre-trained Transformer (OPT)~\cite{zhang2022opt} network as our backbone. We also adopt the tokenizer $\mathcal{T}$ from MeshAnythingV2 to transform a mesh into a sequence $S=\mathcal{T}(\mathcal{M})$. 
The sequence $S$ is defined as an ordered list consisting of control tokens (<split>, <start>, <end>) and vertex coordinate tokens $V$.
The <start> and <end> tokens indicate the sequence bounds.
The <split> token splits the sequence into subsequences containing only vertex tokens, which define a series of adjacent triangles~\cite{chen2024meshanythingv2artistcreatedmesh}. 
{Please refer to the supplemental for more sequence structure details.
}

To perform mesh addition, we use the OPT model to autoregressively predict the probability distribution $P$ of the next sequence token $S_r^{\prime(i+1)}$ given the sketch $\mathcal{I}$, input mesh sequence $S_k=\mathcal{T}(\mathcal{M}_k)$, and the previously generated tokens $S_r^{\prime(1\ldots i)}$:
\begin{equation}
    P\left(S_r^{\prime(i+1)}|S_k,\mathcal{I},S_r^{\prime(1\ldots i)}\right) = \text{OPT}\left(S_k,\mathcal{I},S_r^{\prime(1\ldots i)}\right).
\end{equation}
The token $S_r^{\prime(i+1)}$ is obtained by sampling with the distribution $P$, and then conditions the generation of $S_r^{\prime(i+2)}$.
We detokenize the generated sequence $S^\prime_r$ to mesh triangles and then merge them with the input partial mesh:
\begin{equation}
    \mathcal{M}^\prime = \mathcal{M}_k \cup \mathcal{T}^{-1}(S^\prime_r).
\end{equation}
Instead of generating one token at a time, we introduce a vertex-aligned speculator to predict multiple tokens in a single run (\autoref{sec:speculator}).

The deletion network predicts a mask label for each vertex with a deletion head, which processes the encoded states of every three coordinate tokens corresponding to a vertex. While using the same OPT architecture, the deletion network does not share weights with the addition network. The attention layers in the deletion network are switched to perform bi-directional attention for each position to capture the global context. After the inference, we aggregate $\mathcal{V}^\prime_r$ according to \autoref{eq:binary_deletion_label}, and then obtain the deleted mesh with \autoref{eq:deletion}.

\subsection{Vertex-Aligned Speculator}
\label{sec:speculator}

As we perform mesh addition autoregressively, generating an $n$-token sequence requires $n$ forward passes, significantly limiting mesh generation speed for interactive applications.
One effective way of accelerating generation is to use a speculator to generate multiple tokens in a single run. In contrast to natural language, a tokenized mesh sequence contains strong low-level structure---each vertex $\boldsymbol{v}$ is defined by exactly $3$ tokens $V_{\{x,y,z\}}$ representing its $x$, $y$, and $z$ coordinates. Aligning the speculator with vertex tokens by letting it predict only $V_{\{y,z\}}$, as shown in \autoref{fig:network}, ensures a consistent context of input and output for the speculator. 

We adopt the MLP speculator described in~\cite{wertheimer2024speculator}. In natural language processing tasks, a speculator can be trained with a frozen pre-trained large language model to accelerate generation. However, in our experiments, we found that this results in performance degradation. 
We thus train our speculator jointly with the the mesh addition task, so that the addition transformer learns to inform the speculator with contextual information in hidden states.

As shown in \autoref{fig:network}, the speculator predicts the $y$, $z$ coordinates of a vertex $\boldsymbol{v}^\prime$ by:
\begin{equation}
    P\left(V^\prime_{\{y,z\}}\right) = \text{Speculator}\left(E_x, V^\prime_x\right),
\end{equation}
where $E_x$ represents Transformer hidden states corresponding to $V_x$. 
The speculator is trained with a cross-entropy loss between ground truth $V_{\{y,z\}}$ and the prediction. When jointly trained with OPT, the OPT loss function supervises $V_x$ and control tokens only.

\subsection{Training Data Generation for Mesh Editing}
\label{sec:data_processing}
\begin{figure}[t]
    \centering
    \includegraphics[width=\linewidth]{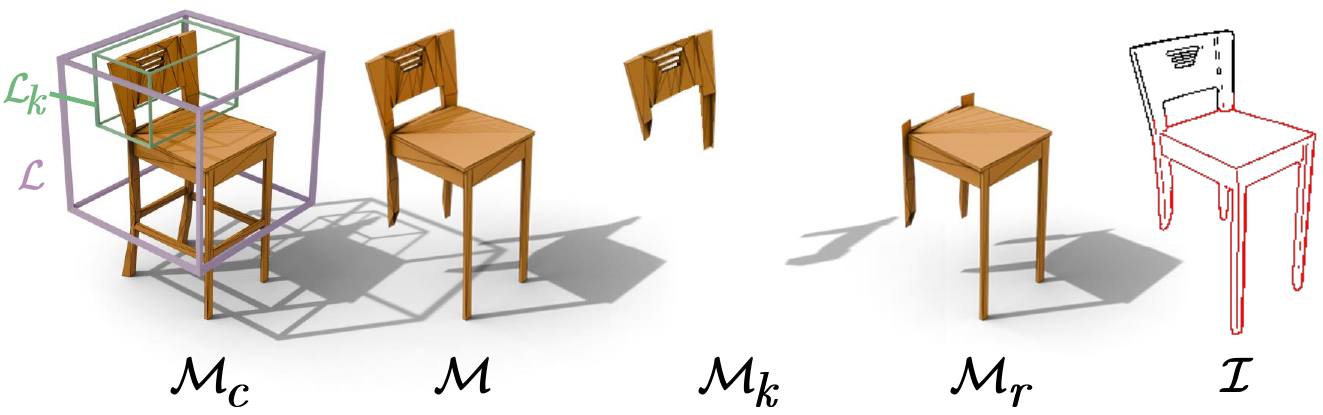}
    \vspace{-0.75cm}
    \caption{An example of training data generation: the volume $\mathcal{L}$ is sampled to cover a large portion from the top, while $\mathcal{L}_k$ covers the chair backrest. The meshes $\mathcal{M}$ and $\mathcal{M}_k$ contain triangles with any vertex in $\mathcal{L}$ and $\mathcal{L}_k$, respectively---that is, triangles fully or partially inside the volumes. $\mathcal{M}_r$ is the difference between $\mathcal{M}$ and $\mathcal{M}_k$. The red and black line strokes in sketch $\mathcal{I}$ corresponds to $\mathcal{M}_k$, and $\mathcal{M}_r$, respectively. \vspace{-0.4cm}}
    \label{fig:data_sample}
\end{figure}

By decomposing sketch-conditioned mesh editing into separate deletion and addition subtasks, we can generate training data for each subtask  
from 3D shape datasets without requiring real mesh editing sequences. \autoref{fig:data_sample} shows our data generation process.

Given a complete 3D mesh sample $\mathcal{M}_c$, we select its triangles to form two mutually exclusive subsets $\mathcal{M}_r$ and $\mathcal{M}_k$, representing the edited and unedited parts, respectively. Note that $\mathcal{M}_r \cup \mathcal{M}_k = \mathcal{M}_c$ is not required, as intermediate meshes in interactive editing can be incomplete before deletion or after addition.
To achieve this, we sample two volumes $\mathcal{L}$ and $\mathcal{L}_k\subsetneq\mathcal{L}$ within the bounding volume of $\mathcal{M}_c$. An example is shown in \autoref{fig:data_sample}.
For each volume, we crop a partial mesh by selecting all triangles containing at least one vertex within the given volume: 
\begin{equation}
    \mathcal{M} = \left\{\mathcal{F}\in\mathcal{M}_c|\exists\boldsymbol{p}\in\mathcal{F}:\boldsymbol{p} \in \mathcal{L}\right\},
\end{equation}
and the same for $\mathcal{M}_k$, $\mathcal{L}_k$.
We compute $\mathcal{M}_r = \mathcal{M} \setminus \mathcal{M}_k$, which is mutually exclusive from $\mathcal{M}_k$.
The volume sampling and mesh cropping details can be found in the supplemental.
We use the resulting meshes to supervise both the addition and the deletion networks defined in \autoref{sec:task}.

\para{Sketch Generation.}
During training, we generate synthetic sketches corresponding to the processed mesh data by performing Canny edge detection on mesh normal and depth renderings. 
{We randomly sample rendering viewpoints and apply image augmentations to synthesize different line drawing styles. The trained model can thus handle human freehand sketches from various viewpoints. We provide the sketch generation details in the supplemental.}

\begin{figure*}[t]
    \centering
    \includegraphics[width=\linewidth]{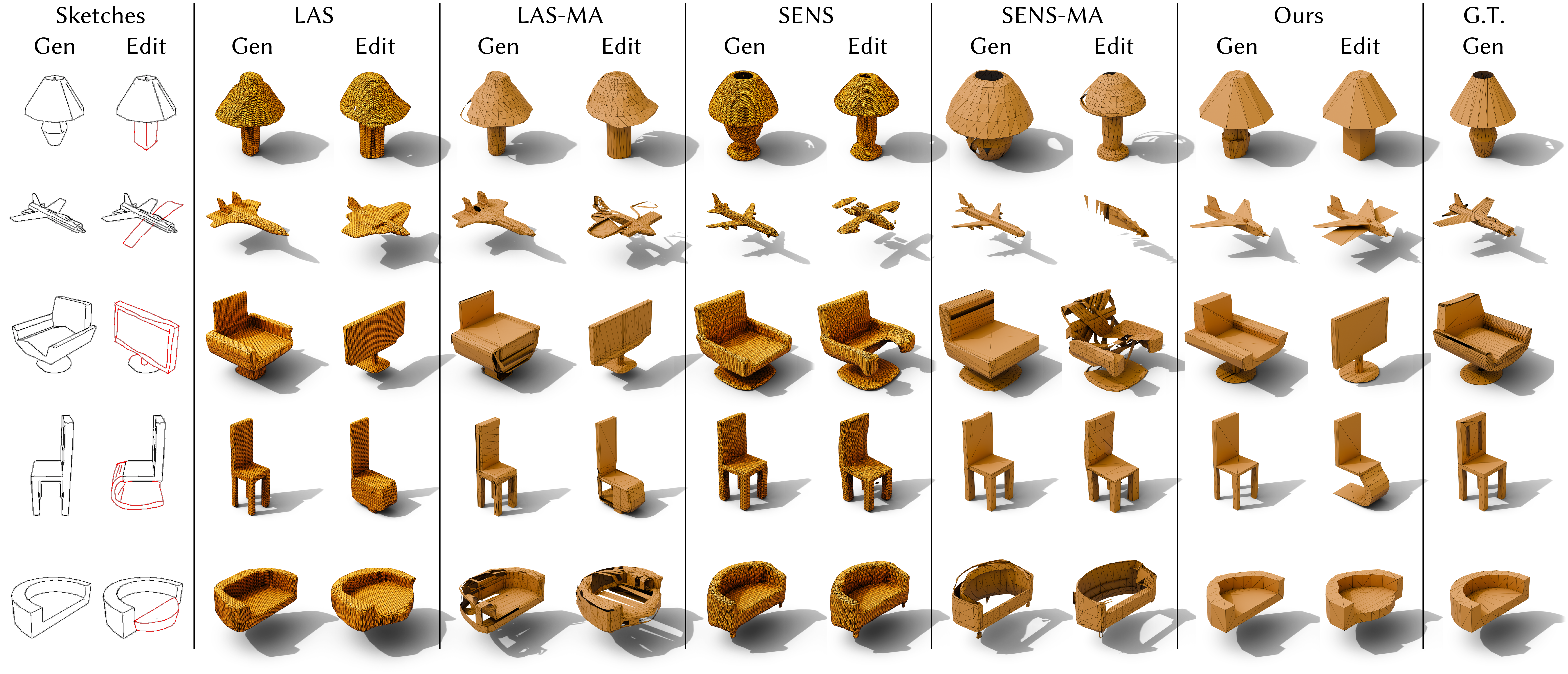}
    \vspace{-0.9cm}
    \caption{Visual comparison of generation and editing results. We show generation and editing results of LAS~\cite{zheng2023lasdiffusion}, SENS~\cite{Binninger:SENS:2024}, and their results post-processed by MeshAnythingV2~(MA)~\cite{chen2024meshanythingv2artistcreatedmesh}. Our method produces 3D meshes that not only  match the sketch inputs and edits, but also offer high fidelity. \vspace{-0.6cm}}
    \label{fig:comparison_table}
\end{figure*}

\section{Experiments}
\label{sec:experiments}

\subsection{Implementation Details}
\label{sec:implementation_details}

\para{Data.} 
Following MeshGPT~\cite{siddiqui2023meshgpt} and MeshAnythingV2~\cite{chen2024meshanythingv2artistcreatedmesh}, we preprocess and filter the ShapeNet dataset~\cite{chang2015shapenet} to obtain approximately $28$k meshes from $55$ categories, which each have  a face count $<768$ for training. We sample around $500$ meshes for validation and $1000$ meshes for testing. Other meshes are used for training. 

\para{Training.}
Both the mesh addition and deletion networks are initialized using the OPT weights from the MeshAnythingV2 checkpoint. The RADIO model is frozen during training. We employ a learning rate schedule from $1\times10^{-4}$ to $1\times10^{-6}$. For the addition model, training begins with a batch size of $56$ for $128$ epochs, followed by $20$ epochs with a batch size of $224$. The deletion model is trained with a batch size of $32$ for $128$ epochs. Using 4 NVIDIA A100 GPUs, the training process takes approximately $5$ days for the addition network and $2$ days for the deletion network.

\subsection{Metrics and Baselines}

We evaluate the generated mesh quality as well as its consistency with the input sketch. For mesh quality, we report the Chamfer Distance~(CD) to the target mesh. A smaller CD indicates better geometric consistency with the target. For perceptual  mesh quality evaluation, we employ shading image-based Fréchet Inception distance~(FID) following MeshGPT~\cite{siddiqui2023meshgpt} and SENS~\cite{Binninger:SENS:2024}. 
We also evaluate sketch-based CLIP~\cite{radford2021clip} similarity, following LAS~\cite{zheng2023lasdiffusion}, to measure the similarity between the generated mesh and the sketch. We additionally introduce sketch-based LPIPS~\cite{zhang2018lpips} to evaluate sketch-to-mesh correspondence: while CLIP shows the similarity of the global context, LPIPS focuses more on local features. For our method, we analyze runtime with the token generation speed in $tokens/second$~(T/s). For this, we measure the time of generating the first $300$ tokens of the addition task, which eliminates the speed difference caused by the varying lengths of  generated sequences. 

We further conduct both unary and binary perceptual studies. In the unary study, 
participants evaluate mesh quality (GQ) and sketch matching (GM) for mesh generation. For mesh editing, they rate edited mesh quality (EQ), edited sketch matching (EM), and edited mesh consistency (EC) of the unedited part.
Each is rated on a scale of $1$ to $5$ ($=$best). In the binary study, we ask participants to compare our method with baselines and select their preferred method. We report the percentages of preferences for our method compared to the baselines.

We compare with two state-of-the-art sketch-conditioned 3D generation methods, LAS~\cite{zheng2023lasdiffusion} and SENS~\cite{Binninger:SENS:2024}. These methods produce signed distance fields or occupancy grids, and we use marching cubes to convert their outputs to triangle meshes. 
{Since LAS and SENS were trained on fewer categories than our method, we report the metrics considering only the results of their respective trained categories (LAS: chair, car, airplane, table, and rifle; SENS: airplane, chair, and lamp). For our method, the metrics are computed on all data in the test set.}
We alternatively use MeshAnythingV2~\cite{chen2024meshanythingv2artistcreatedmesh} to post-process LAS and SENS results (LAS-MA and SENS-MA, respectively), producing more artistically-styled meshes for these baselines. 

\begin{figure*}[t]
    \centering
    \includegraphics[width=\textwidth]{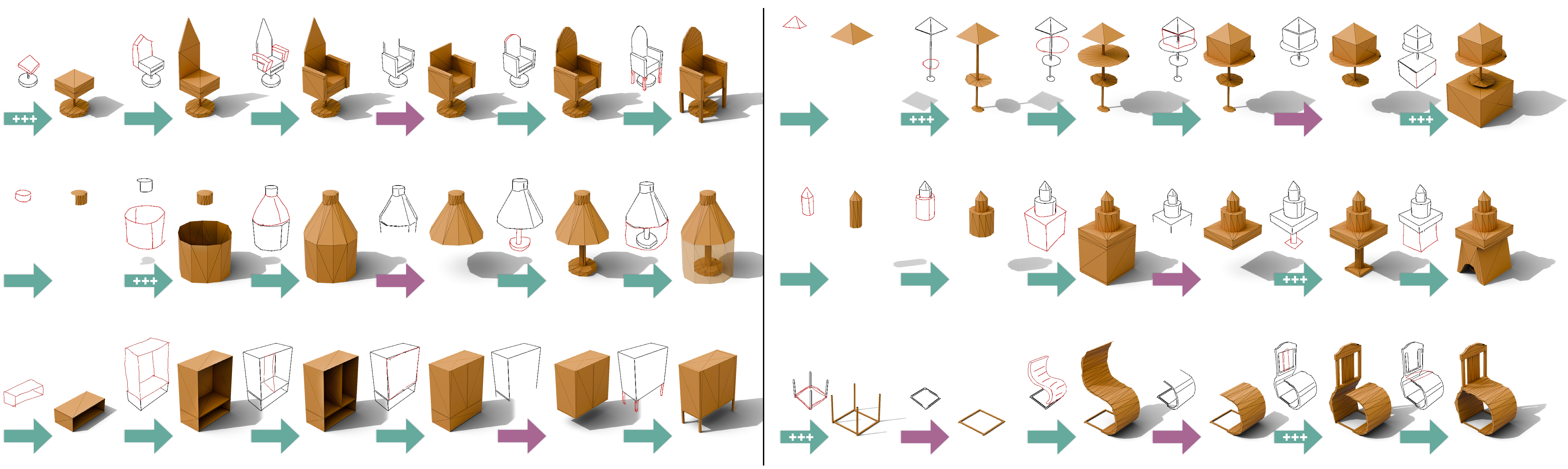}
    \vspace{-0.7cm}
    \caption{Sequence visualization of mesh creation with \name{}. We show the input sketch on the top left of each of our method outputs. \textbf{\textcolor{seq_green}{Green}} and \textbf{\textcolor{seq_purple}{purple}} arrows represent \textbf{\textcolor{seq_green}{addition}} and \textbf{\textcolor{seq_purple}{deletion}}, respectively. The ``+++'' inside a green arrow indicates multiple addition steps. \vspace{-0.8cm}}
    \label{fig:editing_sequence}
\end{figure*}

\subsection{Experimental Setup}

We use the test set described in \autoref{sec:implementation_details} and automatically generate sketches from one of five views: top-left, top-front-left, top-front, top-front-right, or top-right, for the generation task. To further show the robustness of our method, we use the IKEA dataset~\cite{lim2013ikea, sun2018pix3d}, which contains $188$ furniture models. We use automatically generated sketches (following \autoref{sec:data_processing}) for generation tasks because they aid the analysis of sketch-to-mesh correspondence by the CLIP and LPIPS scores. In our qualitative experiments, we show that our method generalizes to hand-drawn sketches.
Additionally, we randomly select $50$ shapes of airplanes, chairs, and lamps (on which SENS is trained) from the two datasets, automatically generate their sketches for generation, and then manually edit the sketches for one-step editing evaluation. 

For the generation task, our method predicts an entire mesh from scratch conditioned on the sketch input. We provide additional camera view conditions in all LAS runs as suggested in \cite{zheng2023lasdiffusion}. We also apply dilation and resizing to sketches to match the input to the training data of LAS. As SENS is category-specific, we use its checkpoint corresponding to the category of the target mesh for each evaluation run. 

For the editing task, we let LAS generate from scratch conditioned on the edited sketch, since it does not support localized editing. SENS was designed for editing, and we ran its editing using its original design. 
For our method, while providing only the original sketch and the edited sketch, we first automatically generate input for deletion by identifying the difference between the two sketches, then perform the deletion-addition operation.

\subsection{Comparison to State of the Art}

\para{Mesh Generation.}  \autoref{tab:generation} evaluates \name{} in comparison with state of the art on mesh generation. Quantitatively, our method outperforms baselines in all evaluation metrics, indicating superior mesh quality and sketch-to-mesh correspondence. We also conduct a perceptual study with $35$ participants, and show results in  Tabs.~\ref{tab:editing} and \ref{tab:user_preference}, further demonstrating user preference for our generated meshes.
\autoref{fig:comparison_table} provides qualitative comparison across the methods. While MeshAnythingV2 can convert marching cube meshes to meshes with more artist-reminiscent design, it is not robust enough to handle various artifacts in outputs of LAS and SENS (e.g., floaters or incomplete parts), resulting in an overall worse quality than our method, which bridges directly sketch to  mesh.

\begin{table}[t]
    \centering
    \resizebox{1.0\columnwidth}{!}{%
        \begin{tabular}{l l c c c c}
            \hline
            \textbf{Dataset} & \textbf{Method} & \textbf{CD}$\downarrow$ & \textbf{LPIPS}$\downarrow$ & \textbf{CLIP}$\uparrow$ & \textbf{FID}$\downarrow$ \\
            \hline
            \multirow{5}{*}{ShapeNet}
                             & LAS             & 15.02                   & 0.2742                     & 94.61                   & 46.90                    \\
                             & LAS-MA          & 22.06                   & 0.2963                     & 93.63                   & \second{18.52}           \\
                             & SENS            & \second{8.95}           & \second{0.2753}            & \second{93.36}          & 81.88                    \\
                             & SENS-MA         & 29.43                   & 0.3348                     & 91.88                   & 42.93                    \\
                             & Ours            & \best{6.20}             & \best{0.1790}              & \best{95.85}            & \best{9.38}              \\
            \hline
            \multirow{5}{*}{IKEA}
                             & LAS             & 16.27                   & 0.2970                     & 94.83                   & 69.54                    \\
                             & LAS-MA          & 20.36                   & 0.3151                     & 93.79                   & \second{46.56}           \\
                             & SENS            & \second{8.76}           & \second{0.2722}            & \second{94.38}          & 115.33                   \\
                             & SENS-MA         & 19.18                   & 0.3037                     & 94.37                   & 99.08                    \\
                             & Ours            & \best{6.78}             & \best{0.1837}              & \best{96.67}            & \best{29.67}             \\
            \hline
        \end{tabular}
    }
    \vspace{-0.3cm}
    \caption{Evaluation on sketch-conditioned mesh generation. We compare with LAS~\cite{zheng2023lasdiffusion}, SENS~\cite{Binninger:SENS:2024}, and further use MeshAnythingV2~(MA)~\cite{chen2024meshanythingv2artistcreatedmesh} to post-process LAS and SENS outputs. The unit of CD is 0.001. Our results are generated by one addition operation from scratch. \name{} outperforms baselines in both mesh quality and consistency with input sketches. \vspace{-0.4cm}}
    \label{tab:generation}
\end{table}

\begin{table}[t]
    \centering
    \resizebox{1.0\columnwidth}{!}{%
        \begin{tabular}{l | cc | cc | ccc}
            \hline
             & \multicolumn{2}{c|}{Sketch Matching} & \multicolumn{2}{c|}{Gen. Rating} & \multicolumn{3}{c}{Edit Rating} \\
            \textbf{Method} & \textbf{LPIPS}$\downarrow$ & \textbf{CLIP}$\uparrow$ & \textbf{GQ}$\uparrow$                            & \textbf{GM}$\uparrow$ & \textbf{EQ}$\uparrow$ & \textbf{EM}$\uparrow$ & \textbf{EC}$\uparrow$ \\
            \hline
            LAS             & 0.3164                     & \second{92.73}
            & 3.2 
                            & 3.0 
                            & 2.7 
                            & 2.2 
                            & 2.5                                                                 \\
            LAS-MA          & 0.3357                     & 91.52
            & 2.8
                            & 2.6 
                            & 2.4 
                            & 2.0 
                            & 2.0                                                                 \\
            SENS            & \second{0.3179}            & 91.49
            & 3.4 
                            & 3.5 
                            & 2.9 
                            & 1.8 
                            & 3.7                                                                 \\
            SENS-MA         & 0.3651                     & 90.20
            & 2.7 
                            & 2.8 
                            & 2.1 
                            & 1.6 
                            & 2.5                                                                 \\
            Ours            & \best{0.2218}              & \best{95.71}
            & \best{4.3}
                            & \best{4.3}
                            & \best{4.3}
                            & \best{4.2}
                            & \best{4.3}                                                         \\
            \hline
        \end{tabular}
    }
    \vspace{-0.3cm}
    \caption{Sketch-based metrics and unary perceptual study ratings (ranged $1$-$5$) on generation and editing results of our hand-drawn sketch evaluation set. We benchmark against LAS~\cite{zheng2023lasdiffusion} and SENS~\cite{Binninger:SENS:2024}, with outputs further refined using MeshAnythingV2 (MA)~\cite{chen2024meshanythingv2artistcreatedmesh}. 
    For mesh generation, participants evaluate mesh quality (GQ) and sketch matching (GM). For mesh editing, participants rate edited mesh quality (EQ), edited sketch matching (EM), and edited mesh consistency (EC) of the unedited part. \vspace{-0.2cm}
    }
    \label{tab:editing}
\end{table}

\begin{table}[t]
    \centering
    \resizebox{0.9\columnwidth}{!}{%
        \begin{tabular}{l c c c c}
            \hline
            \textbf{Operation} & \textbf{LAS} & \textbf{LAS-MA} & \textbf{SENS} & \textbf{SENS-MA} \\
            \hline
            Generation         & 93.3\%       & 94.2\%          & 83.7\%        & 91.0\%           \\
            Editing            & 94.4\%       & 96.5\%          & 91.9\%        & 94.4\%           \\
            \hline
        \end{tabular}
    }
    \vspace{-0.3cm}
    \caption{Binary perceptual study.
   Participants choose their preferred output between our method and each baseline (LAS~\cite{zheng2023lasdiffusion}, SENS~\cite{Binninger:SENS:2024}, and their results post-processed by MeshAnythingV2~(MA)~\cite{chen2024meshanythingv2artistcreatedmesh}). Our method is preferred by a large margin for both generation and editing. \vspace{-0.6cm}}
    \label{tab:user_preference}
\end{table}

\para{Mesh Editing.}  Based on the results of the perceptual study, together with the sketch metrics reported in Tabs. \ref{tab:editing} and \ref{tab:user_preference}, our method achieves the best LPIPS and CLIP scores among baselines and reaches the highest rating and preference among users. From the visual comparisons provided in \autoref{fig:comparison_table}, we notice that baseline methods suffer not only from poor editing results but also from inconsistency in unedited regions. This is because these methods regenerate the whole shape after the edit, can change regions not intended to be edited. Our method solves this problem by decoupling the editing procedure into deletion and addition, ensuring  that only edited regions are changed. 
{We present editing sequences with real human inputs in \autoref{fig:editing_sequence}, demonstrating our method's generalizability. Moreover, our method can edit meshes that are not generated by our networks, which is difficult for baselines, and the results are provided in the supplemental.}

\begin{figure}[t]
    \centering
    \includegraphics[width=0.9\linewidth]{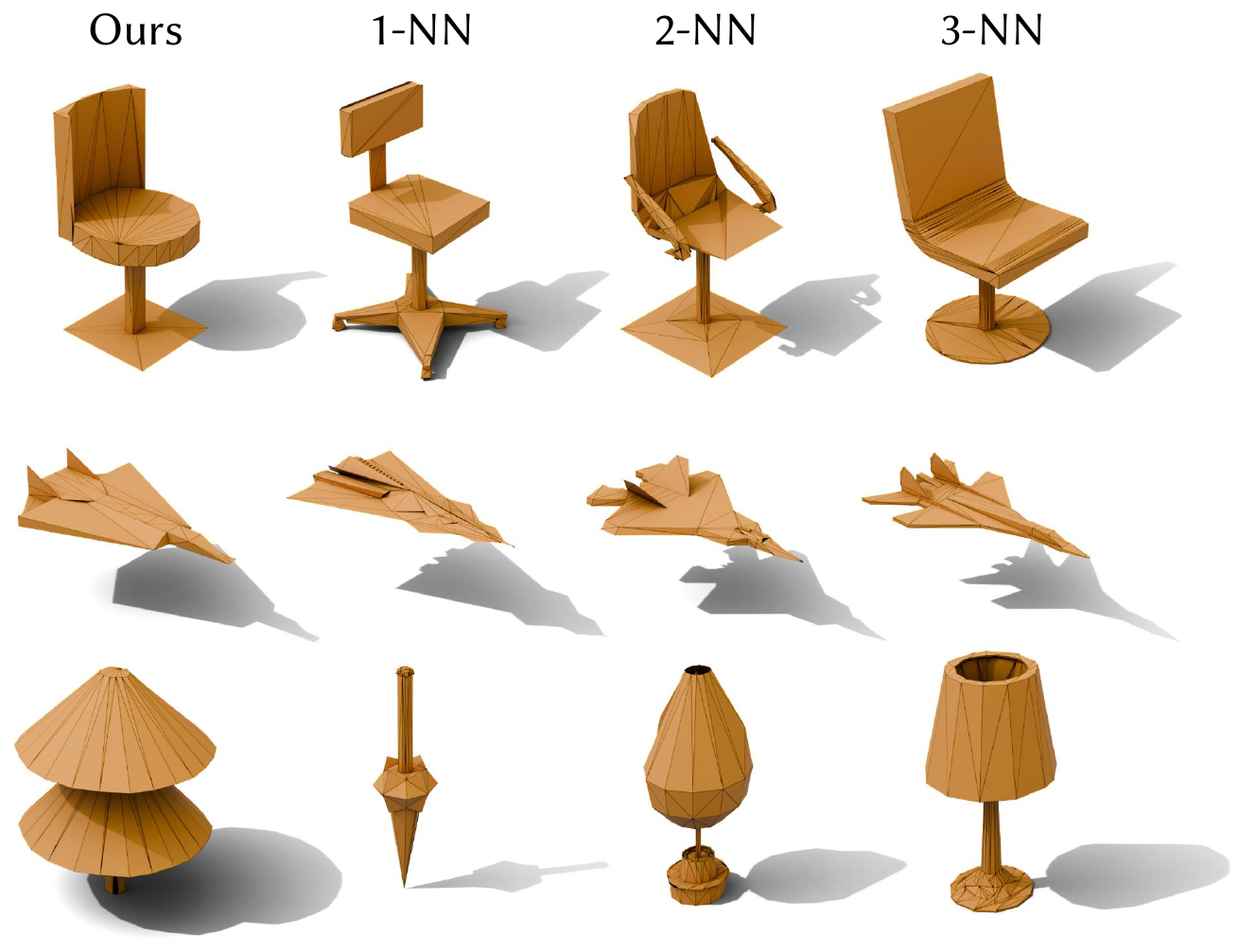}
    \vspace{-0.3cm}
    \caption{Novelty of interactively created shapes. We retrieve the nearest neighbors of the meshes by first aligning them to each train mesh with Iterative Closest Point (ICP) and then computing the Chamfer distance. \name{} can generate new shapes different from those in the train set and can generate shapes beyond the categories of the training data (third row). \vspace{-0.4cm}}
    \label{fig:nn}
\end{figure}

\para{Novelty Analysis.} \autoref{fig:nn} shows a  qualitative novelty analysis, retrieving for our interactively created meshes the 3 nearest neighbors from the training set, based on Chamfer distance. 
Our generated shapes differ from their nearest neighbors in both overall shape and triangulation, which demonstrates that our method can generate new shapes.

\subsection{Ablations}

\autoref{tab:ablation} ablates our vertex-aligned speculator design choices.
We conduct ablation studies on three variants of the addition network: without speculator, without vertex-aligned decoding (with the same decoding length of $2$), and without joint training with the OPT.

\para{Joint training provides context-rich hidden states as speculator input.}
We find that the vanilla speculator training scheme with frozen transformer~\cite{wertheimer2024speculator}
does not work in our mesh generation setting, as the hidden states from our pre-trained transformer do not contain enough contextual information. 
As shown in \autoref{tab:ablation}, by instead training the speculator jointly with the OPT, our model achieves strong mesh quality performance.
In contrast, training the speculator with the OPT frozen results in poor performance. This indicates the OPT learns to encode contexts into the hidden states when jointly trained.

\para{Vertex-alignment preserves mesh quality while achieving comparable speedup as vanilla MLP speculators.}
Comparing no speculator (w/o speculator) and speculator without vertex alignment (w/o vert-alignment), we can see that the speculator, without aligning to vertex tokens, results in a notable speedup, but at large penalty cost in mesh quality and sketch-to-mesh correspondence. By aligning the speculator to vertex tokens, our method reaches the same level of sketch-to-mesh correspondence as the no speculator version while maintaining the naive speculator's advantage in generation speed. Additionally, applying the vertex-aligned speculator results in slightly higher metrics in mesh quality. This is likely due to the vertex-aligned speculator reducing the complexity of the linear prediction head. Rather than predicting control tokens and all $x$, $y$, and $z$ coordinate tokens, it is trained to predict only control tokens and $x$ coordinate tokens, which reduces the overall prediction error. Thus, equipped with the vertex-aligned speculator, our method achieves strong performance and fast generation speeds (approximately $26.7$ faces per second considering the average face density of the mesh sequences). As a result, mesh additions, which typically generate about $\nicefrac{1}{5}$ of a complete shape, run in  about 1 to 5 seconds (depending on the number of faces added), enabling interactive runtimes for editing.

\para{Limitations.} 
{Similar to recent mesh generation methods \cite{siddiqui2023meshgpt, chen2024meshxl}, our transformer model is limited by context length.
Although there are recent attempts~\cite{hao2024meshtron} to extend the context length, their outputs (thousands of triangles) still fall below the requirement of modern computer graphics applications (millions of triangles).
Leveraging techniques, like token compression~\cite{jiang2023llmlingua}, is a promising avenue to explore.
Nevertheless, our approach achieves interactive mesh generation and editing with easy-to-use sketches, enabling iterative creation of novel and complex 3D shapes.}

\begin{table}[t]
    \centering
    \resizebox{1.0\columnwidth}{!}{%
        \begin{tabular}{l | c c c c | c}
            \hline
                & \multicolumn{4}{c|}{Quality Metrics} & Speed \\
            \textbf{Method}    & \textbf{CD}$\downarrow$ & \textbf{LPIPS}$\downarrow$ & \textbf{CLIP}$\uparrow$ & \textbf{FID}$\downarrow$ & \textbf{T/s}$\uparrow$ \\
            \hline
            w/o speculator     & \underline{7.66}           & \best{0.1765}              & \underline{95.59}          & \underline{32.59}           & 60.7                   \\
            w/o vert-alignment      & 9.00                    & 0.1992                     & 94.43                   & 35.65                    & \best{138.9}           \\
            w/o joint training & 57.13                   & 0.5134                     & 84.52                   & 211.46                   & 130.8                  \\
            Ours               & \best{6.78}             & \underline{0.1837}            & \best{96.67}            & \best{29.67}             & \underline{131.1}         \\
            \hline
        \end{tabular}
    }
    \vspace{-0.3cm}
    \caption{Ablation study. We ablate our speculator, vertex alignment of our speculator (using a common MLP speculator~\cite{wertheimer2024speculator} with the same length of $2$), and its joint training with our mesh transformer. The unit of CD is 0.001. T/s represents tokens per second and is measured on NVIDIA A100. Our vertex-aligned speculator accelerates generation by $2.16\times$ without loss in mesh quality. 
    Compared to an MLP speculator without vertex-alignment, which suffers from quality loss, our method runs at a comparable speed.
    \vspace{-0.4cm} }
    \label{tab:ablation}
\end{table}

\section{Conclusion}
\label{sec:conclusion}

\name{} introduces interactive editing into  autoregressive mesh generation, by decomposing editing into addition and deletion operations on sequentialized mesh tokens.
In contrast to existing works which focus on full shape generation, \name{} enables fine-grained and localized control over the artistic generation process.
To enable interactive feedback, our vertex-aligned speculator notably speeds up mesh face generation, running in just a few seconds in response to user sketch updates.
Our sketch-conditioned approach also enables ease of use in creation of complex 3D meshes from novice users.
We believe that \name{} represents a significant step toward democratizing content creation, enabling everyone to create artist-reminiscent 3D meshes.

\section*{Acknowledgements}
This work was funded by AUDI AG.
Lei Li and Angela Dai were supported by the ERC Starting Grant SpatialSem (101076253) and Matthias Nießner by  the ERC Consolidator Grant Gen3D (101171131).
Thanks to Yiwen Chen for guidance on training MeshAnything V2 and Alexandre Binninger for sharing pre-trained weights of SENS.

{
    \small
    \bibliographystyle{ieeenat_fullname}
    \bibliography{main}
}

\clearpage
\appendix
\section*{Appendix}

In this material, we first introduce the details of sequentialized mesh representation in \autoref{sec:mesh_sequence}, followed by details of our self-supervised data processing in \autoref{sec:data_details}. We introduce the implementation of our user interface in \autoref{sec:user_interface}. Then, we showcase four additional qualitative experiments in \autoref{sec:add_qualitative_exp}. Next, we provide more visual evaluation results in \autoref{sec:more_results} and topological statistics of generated meshes in \autoref{sec:topology_of_generated_meshes}, followed by more details and results from our perceptual study in \autoref{sec:perceptual_study_detail}.

\section{Mesh Token Sequence Details}
\label{sec:mesh_sequence}

As stated in \defaultautoref{subsec:arch}{Sec.\ 3.2}, The sequential representation of a mesh consists of vertex coordinate tokens and control tokens where the <split> token separates the full sequence representing the whole mesh into subsequences corresponding to a ``triangle chain'' of adjacent triangles. Each subsequence has $3n$ vertex coordinate tokens where $n$ is the number of vertices contained in the subsequence. Given a subsequence, we first decode a sequence of vertices $\{\boldsymbol{v}_k\}_{k=0}^{n}$: the $x$, $y$, and $z$ position of $\boldsymbol{v}_k$ is obtained from the coordinate tokens at positions $3k$, $3k+1$, and $3k+2$ in the subsequence, respectively. We then get the triangle chain by forming triangles of three adjacent vertices in the vertex sequence:
\begin{equation}
    \{\mathcal{F}\} = \{\{\boldsymbol{v}_k, \boldsymbol{v}_{k-1}, \boldsymbol{v}_{k-2}\}| \forall k: k>=2\}.
\end{equation}
Merging all triangle chains obtained from all subsequences results in the full mesh.
When converting a mesh into sequence, we first sort every triangle in the order of $z-y-x$, then perform a depth-first traversal from the first triangle to find adjacent triangles as the first subsequence. We then subtract the triangles already traversed and start from a new triangle to obtain the next subsequence. We iterate this process until all triangles are traversed.

Our addition network autoregressively generates one new vertex (three coordinate tokens) or the <split> token for each step. 
If the new vertex's index in the current vertex sequence (measured by its distance from the nearest previous <split> token) is $\geq 2$, it forms a triangle with the two preceding vertices.

\section{Data Generation Details}
\label{sec:data_details}

In this section, we provide additional details for our data generation process (\defaultautoref{sec:data_processing}{Sec.\ 3.4}).

\para{Random sampling of volumes.} To sample volumes that satisfy $\mathcal{L}_k\subsetneq\mathcal{L}$, we first randomly define two volumes $\mathcal{L}_{\{a,b\}}$ within the bounding volume (assuming a unit cube) of the complete mesh $\mathcal{M}_c$. For each volume $\tilde{\mathcal{L}}$ of $\mathcal{L}_{\{a,b\}}$, we sample the volume as following: first, randomly choose an axis $i$ from $\{1,2,3\}$ indicating the $x$, $y$, or $z$ axis, then randomly select one region $\mathcal{R}$ from the four candidates: $[-\infty, a]$, $[a, +\infty]$, $[b-c, b+c]$, $[-\infty, b-c] \cup [b+c, +\infty]$, where $a\in[0.2, 0.8]$, $b\in[0.4,0.6]$ and $c\in[0.1,0.4]$ are uniformly sampled within their defined range. The sampled volume is then defined as:
\begin{equation}
    \tilde{\mathcal{L}} = \{\boldsymbol{p}\in \mathbb{R}^3|p_i\in\mathcal{R}\}.
\end{equation}
Then we define:
\begin{equation}
    \mathcal{L}_k=\mathcal{L}_a;\quad\mathcal{L} = \mathcal{L}_a \cup \mathcal{L}_b,
    \label{eq:volume_def}
\end{equation}
which ensures that $\mathcal{L}_k\subseteq\mathcal{L}$. 

After sampling both volumes, it can happen that $\mathcal{L}_b \subseteq \mathcal{L}_a$, and as a result, $\mathcal{L}=\mathcal{L}_k$ and thus the editing part $\mathcal{M}_r$ is empty. To avoid this, we manually set $\mathcal{L}_b=\mathbb{R}^3$ in this case so that it contains the entire mesh (that is, a mesh completion task for addition). 

The random volume sampling, while effective, can produce data samples that are not well-suited for training. For instance, the target mesh $\mathcal{M}$ for addition could be close to the complete mesh $\mathcal{M}_c$ with just a few triangles missing, and that the sketch is also close to the full sketch of $\mathcal{M}_c$. As a consequence, the model would learn to generate a mesh with missing triangles, even when the user provides a complete sketch. To address this issue, we additionally check the coverage of the visibility mask of $\mathcal{M}$ to the complete mesh's ($\mathcal{M}_c$) and set $\mathcal{L}_b=\mathbb{R}^3$ if the coverage is greater than $95\%$.

\para{Sketch generation.} 
Automatic sketch generation is done by applying Canny edge detection on the rendered depth and normal images of $\mathcal{M}$.
We then merge the two edge detection results to obtain the corresponding synthetic sketch used for training. 
To get mutually exclusive sets of line strokes $\mathcal{I}_{\{r,k\}}$, which correspond to different parts of the mesh,
we further render a visibility mask of $\mathcal{M}_r$ in the same camera view and collect all line strokes within the mask as $\mathcal{I}_{r}$. 
$\mathcal{I}_{k}$ is then defined as the line strokes not in $\mathcal{I}_{r}$. 
In practice, the line strokes from the Canny edge detector could lie outside the mask by $1$ pixel. Therefore, we dilate the visibility mask by $1$ pixel before collecting $\mathcal{I}_{r}$. 
During training, we randomly sample camera views from the upper unit viewing hemisphere, with azimuth angles in  $[-90^\circ, 90^\circ]$ 
and elevation angles in $[0^\circ, 60^\circ]$.

\para{Mesh and sketch augmentations.} During training, we apply random axis independent scaling to the mesh, with a scaling factor uniformly sampled in $[0.9,1.1]$ on each axis. For the sketch, we apply a random affine transformation and a random elastic transformation to fill the domain gap between generated sketches and human drawings.

\begin{figure}[t]
    \centering
    \includegraphics[width=\linewidth]{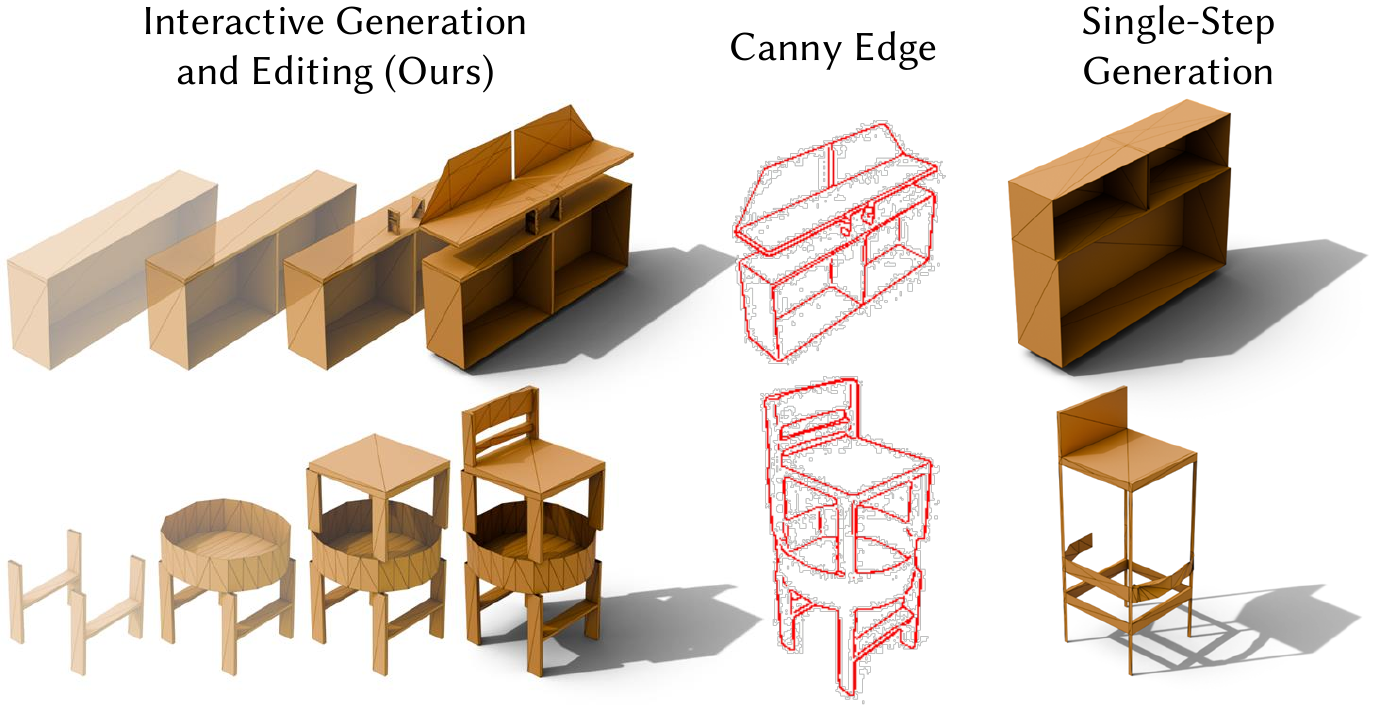}
    \vspace{-0.6cm}
    \caption{ Comparisons between our interactive generation and editing vs. single-step generation.
    \textbf{Left}: meshes interactively created using our method, which effectively uses multiple localized generation steps to achieve high complexity. \textbf{Middle}: synthetic sketches generated from the final meshes via Canny Edge detection. \textbf{Right}: meshes generated by applying one pass of the addition network (without iterative editing) on the middle sketches. Our interactive editing pipeline enables creating intricate shapes that can be challenging to generate in one step.\vspace{-0.4cm}}
    \label{fig:regen}
\end{figure}

\section{User Interface Implementation}
\label{sec:user_interface}
We develop a Gradio~\cite{abid2019gradio} user interface to demonstrate our method in an interactive editing environment, accessible from browsers on all types of devices (PC, iPad, etc.). As visualized in \defaultautoref{fig:teaser}{Fig.\ 1}, the interface contains a sketchpad for users to draw (addition) and erase (deletion) line strokes and a three.js~\cite{threejs} viewer displaying the mesh in real-time. Users can submit their sketches to the addition or deletion job and can always re-sketch and re-submit the job when the outcome should be iterated on. When an edit is accepted, the interface refreshes the sketch in the sketchpad to match the current mesh for further edit, which corresponds to the process depicted in \defaultautoref{fig:overview}{Fig.\ 2}. We refer to our supplementary video for live demos.

\section{Additional Qualitative Experiments}
\label{sec:add_qualitative_exp}

\para{Generating complex shapes through interactive editing.} 
Our interactive mesh editing enables not only an iterative creation process, but its localized editing focus enables construction of complex 3D shapes by decomposing their generation into a sequence of simpler part generations. 
As a result, artists can use our method to create a larger variety of complex shapes than the single-step generation, as demonstrated in \autoref{fig:regen}. We refer to our supplementary  video for more interactive modelings in action.

\begin{figure}[tbp]
    \centering
    \includegraphics[width=0.9\linewidth]{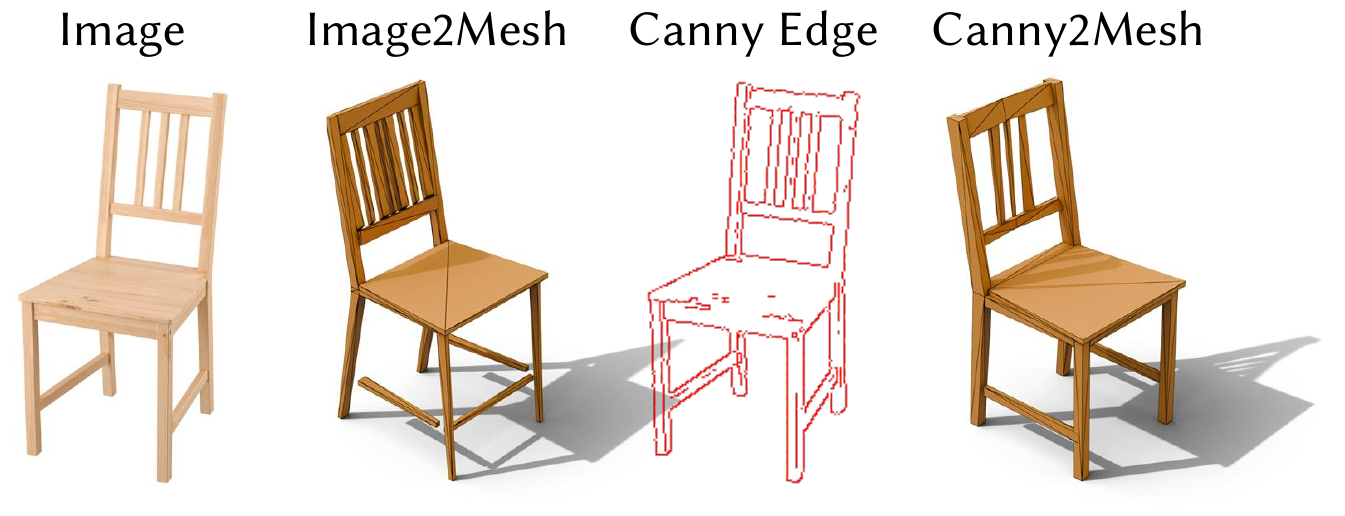}
    \caption{Image to mesh results. Benefiting from the large image foundation model RADIO, our method generalizes to image-conditioning without additional training. For better image matching and mesh quality, users can pre-process the image to get a canny edge for conditioning.\vspace{-0.4cm}}
    \label{fig:image2mesh}
\end{figure}

\para{Image conditioned mesh generation.} 
While our method is not trained on images, it generalizes well to image-conditioning, taking advantage of the large image foundation model RADIO~\cite{Ranzinger2024radio}. It requires no re-training nor any adaptations to the pipeline to use our model for image-conditioned mesh generation. In \autoref{fig:image2mesh}, we show an example of image-conditioned generation with our method. It is worth mentioning that the mesh quality and image-to-mesh correspondence are lower for image-conditioned generation compared to sketch-conditioned generation. A simple pre-processing to get the canny edge for conditioning would boost the performance of our method. 

\begin{figure}[tbp]
    \centering
    \includegraphics[width=0.9\linewidth]{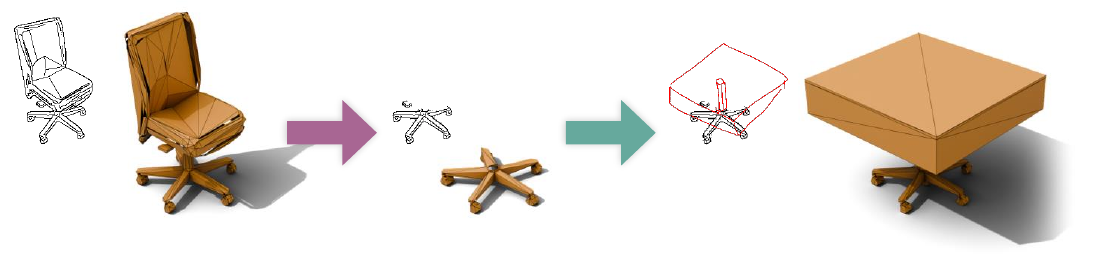}
    \caption{Our method allows us to edit directly on a given mesh (created manually or with other methods). We first load a mesh of an office chair, and automatically generate the corresponding rendered sketch. We can then delete the upper part and add a square tabletop to build a movable table.\vspace{-0.4cm}}
    \label{fig:direct_edit}
\end{figure}
\begin{figure}[b]
    \centering
    \includegraphics[width=\linewidth]{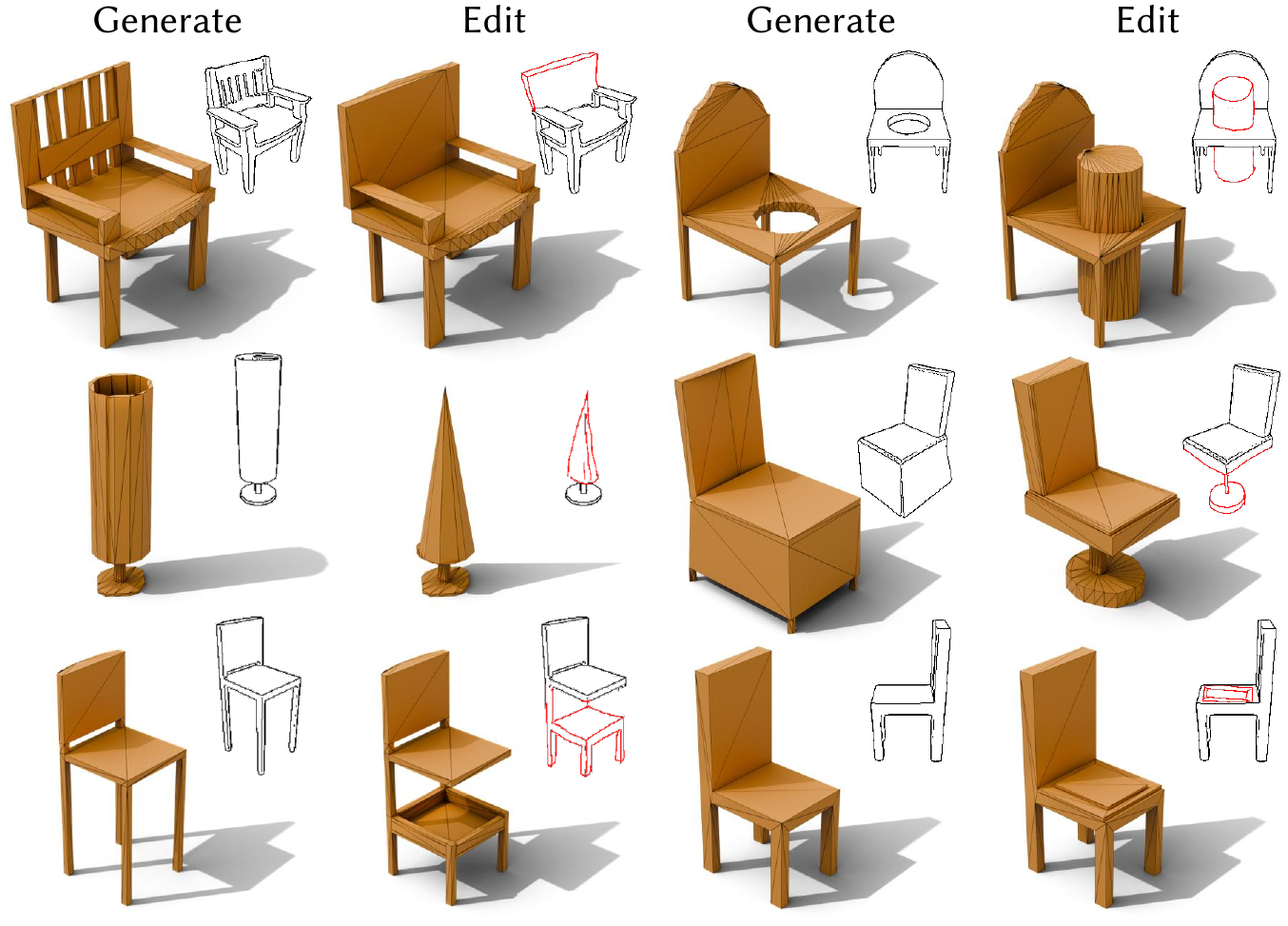}
    \caption{Visual results of our method on our hand-drawn dataset. Our method successfully performs a variety of edits, from local changes to edits that affect the majority of the shape.}
    \label{fig:more_results}
\end{figure}

\para{Direct mesh editing.}
Another benefit of using mesh as an underlying representation is that we could edit directly on a given mesh. Other methods, such as SENS~\cite{Binninger:SENS:2024}, that perform editing on latent spaces struggle at editing a given mesh as it is tricky to convert the mesh to the model input. The common way is to encode the mesh to the latent space and feed it to the model. Consequently, there is no guarantee that the unedited part will remain unchanged due to the error of encoding and decoding. Our method naturally addresses this issue using explicit mesh representation as model input. This allows us to edit a given mesh without worrying about changing the unedited parts. In \autoref{fig:direct_edit}, we show an example of editing a given mesh. After loading the mesh and generating the sketch, we could perform addition and deletion as if the mesh is generated by the model.

\begin{figure}[t]
    \centering
    \includegraphics[width=1.0\linewidth]{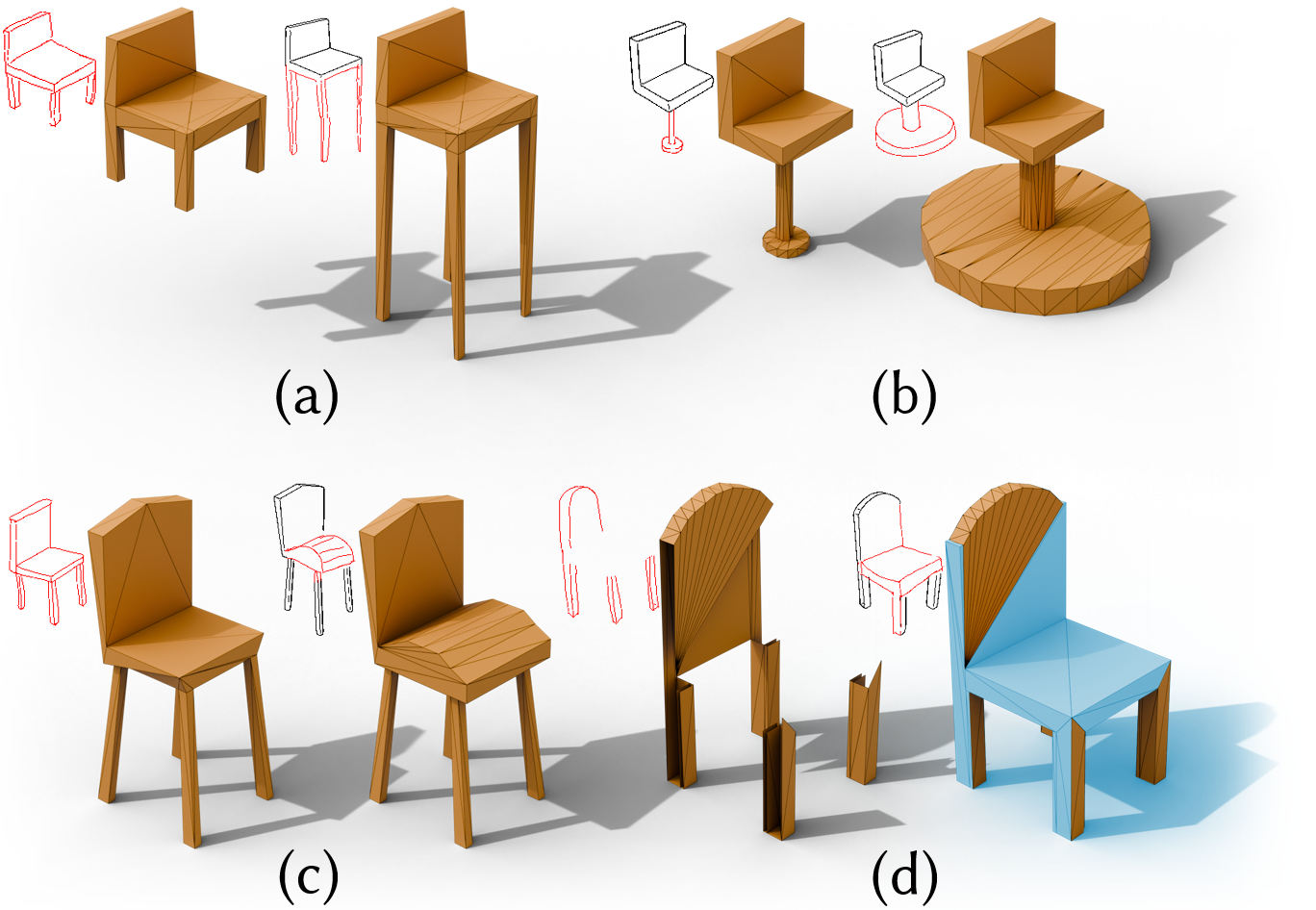}
    \vspace{-0.5cm}
    \caption{Qualitative examples showing the editing capability of our model. 
    \textbf{(a, b)}: sensitivity of edited shapes to the sketch input. \textbf{(c)}: editing the middle part of the shape. \textbf{(d)}: the added part (cyan) connects to the existing part.
    \vspace{-0.3cm}}
    \label{fig:edit_cap}
\end{figure}

\para{Mesh-to-Sketch Alignment.}
Our method achieves superior performance on sketch alignment. \autoref{fig:edit_cap} shows that our method can \textbf{(a, b)}: generate shapes aligning with the length and size of the sketch input; \textbf{(c)}: edit the middle part of the mesh; \textbf{(d)}: complete broken shapes via addition. 
While our method achieves better sketch alignment metrics than baseline methods, it does not guarantee exact alignment between generated meshes and sketches, since our model aims at high-fidelity shape generation from \textit{freehand} sketches.

\begin{figure}[t]
    \centering
    \includegraphics[width=\linewidth]{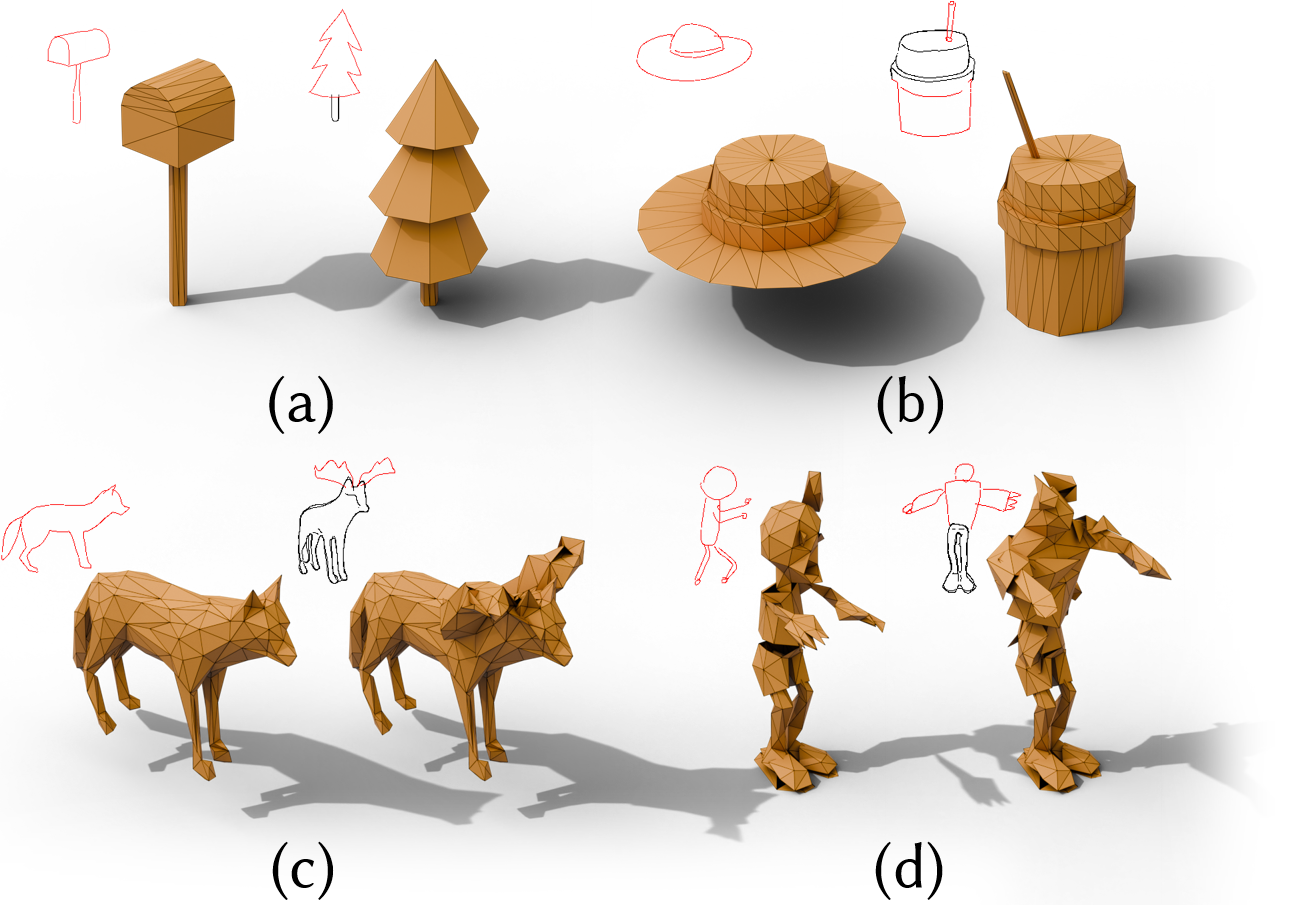}
    \caption{Qualitative results on Objaverse~\cite{objaverse}~\textbf{(a, b)} and DeformingThings4D~\cite{li20214dcomplete}~\textbf{(c, d)}. We fine-tune our networks on the two datasets and show their scalability to multiple categories.}
    \label{fig:other_datasets}
\end{figure}

\section{Additional Visual Results}
\label{sec:more_results}

We provide more mesh generation and editing results of our method in \autoref{fig:more_results}, as a complement to \defaultautoref{fig:comparison_table}{Fig.\ 7}. Note that our hand-drawn sketch editing dataset, while containing only 50 meshes, covers a large variety of editing tasks that are common in interactive mesh editing, from small local changes to global shape structure adjustments. Thus, it serves as a strong and challenging benchmark for sketch-based mesh editing.
While trained only on self-supervised data (\defaultautoref{sec:data_processing}{Sec.\ 3.4}) without any data from real artist drawings, our method generalizes to these tasks proposed with human drawing, showing consistent performance across diverse editing tasks. This highlights the effectiveness of our self-supervised data processing approach.

In \autoref{fig:other_datasets}, we show the mesh editing results of our networks fine-tuned on Objaverse~\cite{objaverse} and DeformingThings4D~\cite{li20214dcomplete}. For Objaverse fine-tuning, we filter the meshes with fewer than $768$ triangles to obtain around $9,000$ meshes. For DeformingThings4D, we sample around $1,000$ meshes from all animations and decimate the mesh to fit the limit of our setup. The result shows the scalability of our method to  a large variety of meshes.

\begin{table}[tb]
\centering
\resizebox{\columnwidth}{!}{%
\begin{tabular}{lcccc}
\toprule
\textbf{Method} & \textbf{\#Faces} & \textbf{\#Components} & \textbf{Non-manifold Edges\%} & \textbf{\#Self Intersections} \\
\midrule
LAS          & 33,573 & 1.0   & 0 & 0 \\
LAS-MA       & 1,114  & 51.9  & 2.05 & 1,738 \\
SENS         & 33,859 & 2.5   & 0 & 0 \\
SENS-MA      & 1,513  & 73.4  & 2.22 & 1,787 \\
Ours         & 231    & 12.3  & 2.01 & 259 \\
Training Data           & 285    & 12.0  & 2.13 & 3,285 \\
\bottomrule
\end{tabular}
}
\vspace{-0.25cm}
\caption{
Topological statistics of edited meshes vs.\ train data. 
Our method generates meshes that align best with the training data with significantly reduced self-intersections. Note that LAS and SENS generate over-tessellated meshes that deviate from the training data.
}
\label{tab:mesh_stats}
\end{table}

\begin{table}[b]
    \centering
    \resizebox{\columnwidth}{!}{%
        \begin{tabular}{l | cc | ccc}
            \hline
             & \multicolumn{2}{c|}{Generation Rating} & \multicolumn{3}{c}{Editing Rating} \\
            \textbf{Method} & \textbf{GQ}$\uparrow$                            & \textbf{GM}$\uparrow$ & \textbf{EQ}$\uparrow$ & \textbf{EM}$\uparrow$ & \textbf{EC}$\uparrow$ \\
            \hline
            LAS             
            & 3.2 \newline \scriptsize{(1.1)}
                            & 3.0 \newline \scriptsize{(1.2)}
                            & 2.7 \newline \scriptsize{(1.1)}
                            & 2.2 \newline \scriptsize{(1.1)}
                            & 2.5 \newline \scriptsize{(1.3)}                                                                \\
            LAS-MA
            & 2.8 \newline \scriptsize{(1.2)}
                            & 2.6 \newline \scriptsize{(1.2)}
                            & 2.4 \newline \scriptsize{(1.3)}
                            & 2.0 \newline \scriptsize{(1.1)}
                            & 2.0 \newline \scriptsize{(1.1)}                                                                \\
            SENS      
            & 3.4 \newline \scriptsize{(1.2)}
                            & 3.5 \newline \scriptsize{(1.1)}
                            & 2.9 \newline \scriptsize{(1.2)}
                            & 1.8 \newline \scriptsize{(1.0)}
                            & 3.7 \newline \scriptsize{(1.6)}                                                                \\
            SENS-MA     
            & 2.7 \newline \scriptsize{(1.4)}
                            & 2.8 \newline \scriptsize{(1.4)}
                            & 2.1 \newline \scriptsize{(1.4)}
                            & 1.6 \newline \scriptsize{(1.1)}
                            & 2.5 \newline \scriptsize{(1.3)}                                                                \\
            Ours       
            & \best{4.3 \newline \scriptsize{(0.9)}}
                            & \best{4.3 \newline \scriptsize{(0.8)}}
                            & \best{4.3 \newline \scriptsize{(0.8)}}
                            & \best{4.2 \newline \scriptsize{(0.9)}}
                            & \best{4.3 \newline \scriptsize{(1.0)}}                                                         \\
            \hline
        \end{tabular}
        }
        \caption{User ratings (ranged $1$-$5$) on generation and editing results of our hand-drawn sketch evaluation set. We benchmark against LAS~\cite{zheng2023lasdiffusion} and SENS~\cite{Binninger:SENS:2024}, with outputs further refined using MeshAnythingV2 (MA)~\cite{chen2024meshanythingv2artistcreatedmesh} as a post-processing baseline. For mesh generation, participants evaluate mesh quality (GQ) and sketch matching (GM). For mesh editing, participants rate edited mesh quality (EQ), edited sketch matching (EM), and edited mesh consistency (EC) of the unedited part.}
    \label{tab:user_study_std}
\end{table}

\section{Topology of Generated Meshes}
\label{sec:topology_of_generated_meshes}
As shown in \autoref{tab:mesh_stats}, our method generates meshes with similar topological statistics to the training data, while meshes of baseline methods deviate from it. LAS and SENS meshes are over-tessellated marching-cube meshes and are not ideal, even with no non-manifold edges and no self-intersections. We also surprisingly find that the number of self-intersections of our method is significantly less than the training data. We believe the reason is that the model learns to use the simplest way to represent a surface, while in ShapeNet, there are often redundant triangles.

\section{Perceptual Study Details}
\label{sec:perceptual_study_detail}

\begin{figure}[t]
    \centering
    \includegraphics[width=0.95\linewidth]{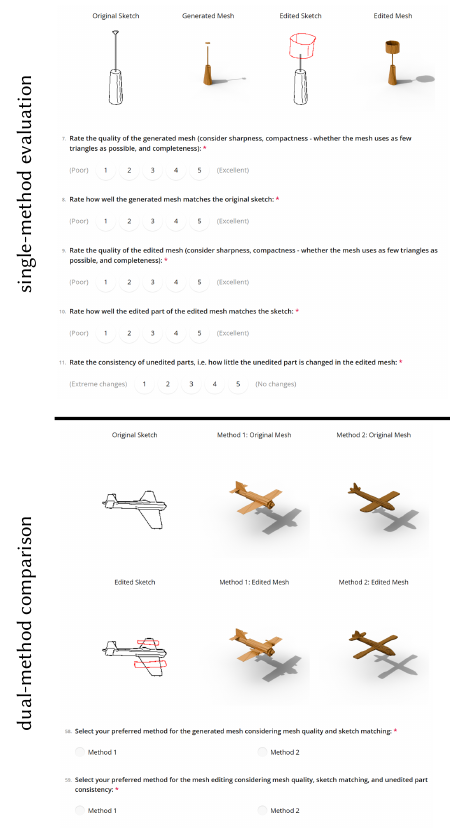}
    \caption{Screenshot of our perceptual study. \textbf{Top}: single-method evaluation. \textbf{Bottom}: dual-method comparison. Each participant is required to complete 10 single-method evaluations and 10 dual-method comparison tasks.\vspace{-0.4cm}}
    \label{fig:userstudy}
\end{figure}

In our perceptual study, we ask each participant to do 10 single-method evaluation tasks and 10 dual-method comparison (ours vs. baseline) tasks. \autoref{fig:userstudy} shows the UI for our perceptual study. The samples used for the questions are randomly chosen from all mesh editing results of all methods and are different for each participant. Our participants have a large variety: from professional artists to hobbyists, students, and individuals with no prior artistic experience.

We show in \autoref{tab:user_study_std} again the unary user rating with the standard deviation. With the smallest deviation observed in all subjects, we show a high agreement among participants regarding the performance of our method.

\end{document}